\documentclass[%
  twocolumn, aps, prb, amsmath, amssymb, superscriptaddress, nofootinbib,
  floatfix ]{revtex4-1}

\usepackage{graphicx}
\usepackage{hyperref}
\usepackage{bm}
\usepackage{color}

\newcommand{\s}{\sum\limits}

\newcommand{\pa}{\partial}

\newcommand{\be}{\begin{equation}} \newcommand{\e}{\end{equation}}
\newcommand{\beml}{\begin{subequations}} \newcommand{\eml}{\end{subequations}}
\newcommand{\beq}{\begin{eqnarray}} \newcommand{\eq}{\end{eqnarray}}
\newcommand{\ba}{\begin{array}} \newcommand{\ea}{\end{array}}
\newcommand{\bpm}{\begin{pmatrix}} \newcommand{\epm}{\end{pmatrix}}
\newcommand{\bc}{\begin{cases}} \newcommand{\ec}{\end{cases}}
\newcommand{\lt}{\left}
\newcommand{\rt}{\right}
\newcommand{\n}{\nonumber}
\newcommand{\la}{\langle}
\newcommand{\ra}{\rangle}
\newcommand{\ep}{\varepsilon}

\newcommand{\bs}{\mathbf}
\newcommand{\bb}{\boldsymbol}

\newcommand{\para}{\parallel}

\renewcommand{\log}{\mathop{\mathrm{ln}}\nolimits}

 \DeclareMathOperator{\tr}{Tr}
 \DeclareMathOperator{\im}{Im}

\begin{document}

\title{Gilbert damping in two-dimensional metallic anti-ferromagnets}

\author{R. J. Sokolewicz}
\affiliation{Radboud University, Institute for Molecules and Materials, 6525 AJ Nijmegen, the Netherlands}
\affiliation{Qblox, Delftechpark 22, 2628 XH Delft, the Netherlands}

\author{M. Baglai}
\affiliation{Department of Physics and Astronomy, Uppsala University, Box 516, SE-751 20, Uppsala, Sweden}

\author{I. A. Ado}
\affiliation{Radboud University, Institute for Molecules and Materials, 6525 AJ Nijmegen, the Netherlands}

\author{M. I. Katsnelson}
\affiliation{Radboud University, Institute for Molecules and Materials, 6525 AJ Nijmegen, the Netherlands}

\author{M. Titov}
\affiliation{Radboud University, Institute for Molecules and Materials, 6525 AJ Nijmegen, the Netherlands}

\date{\today}

\begin{abstract}
A finite spin life-time of conduction electrons may dominate Gilbert damping of
two-dimensional metallic anti-ferromagnets or anti-ferromagnet/metal
heterostructures. We investigate the Gilbert damping tensor for a typical
low-energy model of a metallic anti-ferromagnet system with honeycomb magnetic
lattice and Rashba spin-orbit coupling for conduction electrons. We distinguish
three regimes of spin relaxation: exchange-dominated relaxation for weak
spin-orbit coupling strength, Elliot-Yafet relaxation for moderate spin-orbit
coupling, and Dyakonov-Perel relaxation for strong spin-orbit coupling. We show,
however, that the latter regime takes place only for the in-plane Gilbert
damping component. We also show that anisotropy of Gilbert damping persists for
any finite spin-orbit interaction strength provided we consider no spatial
variation of the N\'eel vector. Isotropic Gilbert damping is restored only if
the electron spin-orbit length is larger than the magnon wavelength. Our theory
applies to MnPS$_3$ monolayer on Pt or to similar systems. 
\end{abstract}

\maketitle

\section{Introduction}

Magnetization dynamics in anti-ferromagnets continue to attract a lot of
attention in the context of possible applications \cite{Hoffmann20, Mazurenko21,
Altermagnetism22,Bernewig22}. Various proposals utilize the possibility of THz
frequency switching of anti-ferromagnetic domains for ultrafast information
storage and computation \cite{Kimel23, Weipeng21}. The rise of van der Waals
magnets has had a further impact on the field due to the possibility of creating
tunable heterostructures that involve anti-ferromagnet and semiconducting layers
\cite{Gibertini2019}. 

Understanding relaxation of both the N\'eel vector and non-equilibrium
magnetization in anti-ferromagnets is recognized to be of great importance for
the functionality of spintronic devices \cite{hals_phenomenology_2011,
cheng_terahertz_2016, urazhdin_effect_2007, cheng_ultrafast_2015,
khymyn_antiferromagnetic_2017, cheng_spin_2014}. On one hand, low Gilbert
damping must generally lead to better electric control of magnetic order via
domain wall motion or ultrafast domain switching \cite{mougin_domain_2007,
thiele_steady-state_1973,weber_gilbert_2019}. On the other hand, an efficient
control of magnetic domains must generally require a strong coupling between
charge and spin degrees of freedom due to a strong spin-orbit interaction, that
is widely thought to be equivalent to strong Gilbert damping. 

In this paper, we focus on a microscopic analysis of Gilbert damping due to
Dyakonov-Perel and Elliot-Yafet mechanisms. We apply the theory to a model of a
two-dimensional N\'eel anti-ferromagnet with a honeycomb magnetic lattice. 

Two-dimensional magnets typically exhibit either easy-plane or easy-axis
anisotropy, and play crucial roles in stabilizing magnetism at finite
temperatures \cite{PhysRevB.60.1082, Spirin2003Magnetization}. Easy-axis
anisotropy selects a specific direction for magnetization, thereby defining an
axis for the magnetic order. In contrast, easy-plane anisotropy does not select
a particular in-plane direction for the Néel vector, allowing it to freely
rotate within the plane. This situation is analogous to the XY model, where the
system's continuous symmetry leads to the suppression of out-of-plane
fluctuations rather than fixing the magnetization in a specific in-plane
direction \cite{kosterlitz1974critical,kosterlitz2018ordering}. Without this
anisotropy, the magnonic fluctuations in a two-dimensional crystal can grow
uncontrollably large to destroy any long-range magnetic order, according to the
Mermin-Wagner theorem \cite{mermin_absence_1966}.

\begin{figure}[bt]
\centerline{
    \includegraphics[width=0.28\linewidth]{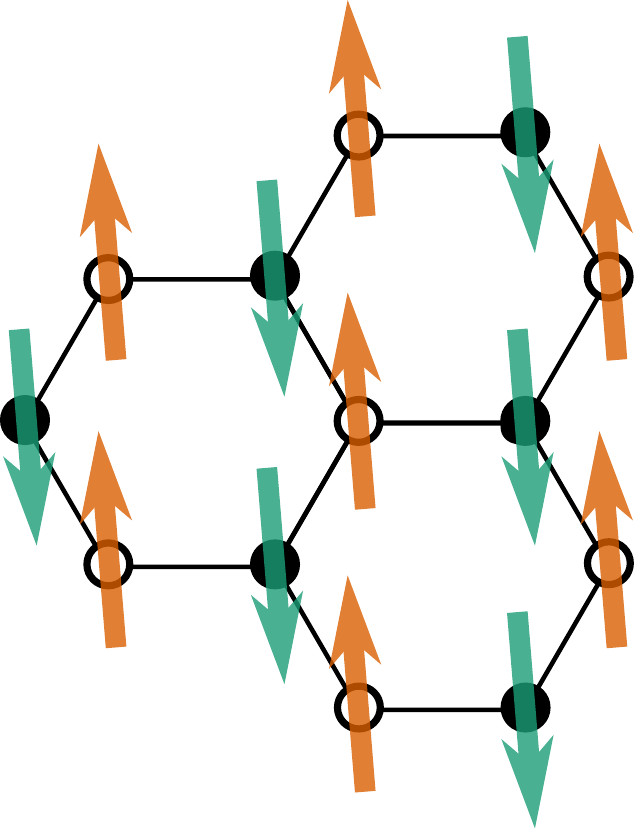}
    \hspace{0.025\textwidth}
    \includegraphics[width=0.28\linewidth]{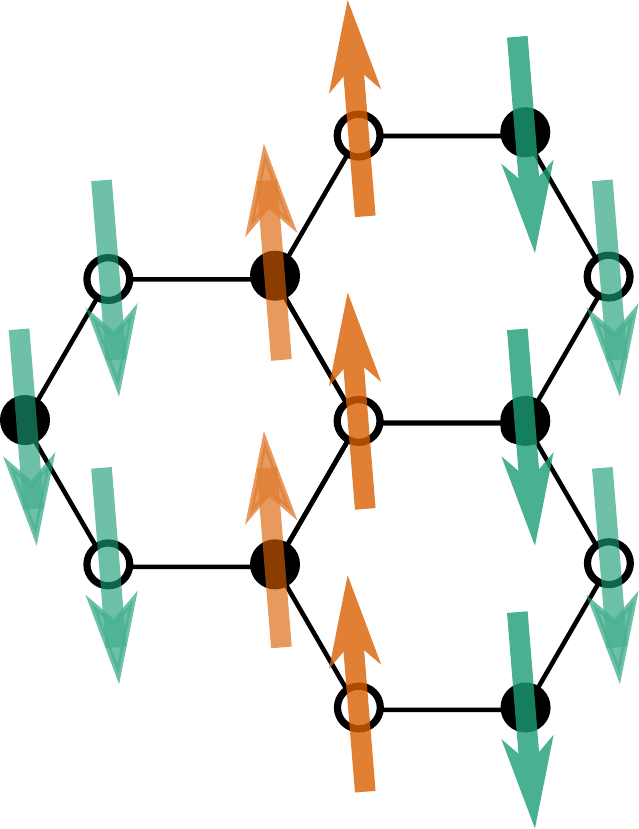}
    \hspace{0.025\textwidth}
    \includegraphics[width=0.28\linewidth]{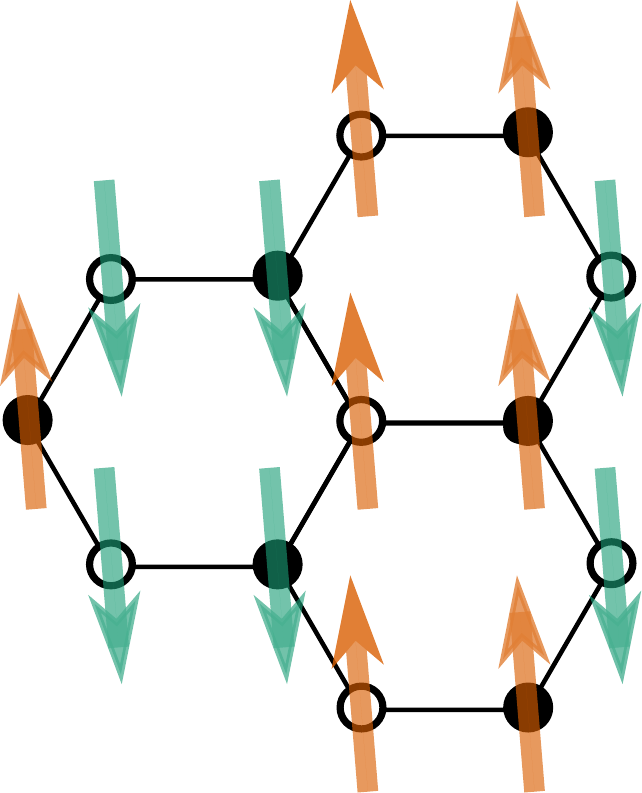}
}
\caption{Three anti-ferromagnetic phases commonly found among van-der-Waals magnets. Left-to-right: N\'eel, zig-zag, and stripy.}
\label{fig:afm_phases}
\end{figure}

Recent density-functional-theory calculations for single-layer transition metal
trichalgenides \cite{chittari_carrier-_2020}, predict the existence of a large
number of metallic anti-ferromagnets with honeycomb lattice and different types
of magnetic order as shown in Fig.~\ref{fig:afm_phases}. Many of these crystals
may have the N\'eel magnetic order as shown in Fig.~\ref{fig:afm_phases}a and
are metallic: FeSiSe$_3$, FeSiTe$_3$, VGeTe$_3$, MnGeS$_3$, FeGeSe$_3$,
FeGeTe$_3$, NiGeSe$_3$, MnSnS$_3$, MnSnS$_3$, MnSnSe$_3$, FeSnSe$_3$, NiSnS$_3$.
Apart from that it has been predicted that anti-ferromagnetism can be induced in
graphene by bringing it in proximity to MnPSe$_3$ \cite{hogl_quantum_2020} or by
bringing it in double proximity between a layer of Cr$_2$Ge$_2$Te$_6$ and WS$_2$
\cite{zollner_purely_2019}.

Partly inspired by these predictions and recent technological advances in
producing single-layer anti-ferromagnet crystals, we propose an effective model
to study spin relaxation in 2D honeycomb anti-ferromagnet with N\'eel magnetic
order. The same system was studied by us in Ref.~\onlinecite{baglai_giant_2020},
where we found that spin-orbit coupling introduces a weak anisotropy in
spin-orbit torque and electric conductivity. Strong spin-orbit coupling was
shown to lead to a giant anisotropy of Gilbert damping. 

Our analysis below is built upon the results of
Ref.~\onlinecite{baglai_giant_2020}, and we investigate and identify three
separate regimes of spin-orbit strength. Each regime is characterized by
qualitatively different dependence of Gilbert damping on spin-orbit interaction
and conduction electron transport time. The regime of weak spin-orbit
interaction is dominated by exchange field relaxation of electron spin, and the
regime of moderate spin-orbit strength is dominated by Elliot-Yafet spin
relaxation. These two regimes are characterized also by a universal factor of 2
anisotropy of Gilbert damping. The regime of strong spin-orbit strength, which
leads to substantial splitting of electron Fermi surfaces, is characterized by
Dyakonov-Perel relaxation of the in-plane spin component and Elliot-Yafet
relaxation of the perpendicular-to-the-plane Gilbert damping which leads to a
giant damping anisotropy. Isotropic Gilbert damping is restored only for finite
magnon wave vectors such that the magnon wavelength is smaller than the
spin-orbit length. 

Gilbert damping in a metallic anti-ferromagnet can be qualitatively understood
in terms of the Fermi surface breathing \cite{fahnle_breathing_2006}. A change
in the magnetization direction gives rise to a change in the Fermi surface to
which the conduction electrons have to adjust. This electronic reconfiguration
is achieved through the scattering of electrons off impurities, during which
angular momentum is transferred to the lattice. Gilbert damping, then, should be
proportional to both (i) the ratio of the spin life-time and momentum life-time
of conduction electrons, and (ii) the electric conductivity. Keeping in mind
that the conductivity itself is proportional to momentum life-time, one may
conclude that the Gilbert damping is linearly proportional to the spin life-time
of conduction electrons. At the same time, the spin life-time of localized spins
is \emph{inversely} proportional to the spin life-time of conduction electrons.
A similar relation between the spin life-times of conduction and localized
electrons also holds for relaxation mechanisms that involve electron-magnon
scattering \cite{simensen_magnon_2020}.

Our approach formally decomposes the magnetic system into a classical sub-system
of localized magnetic moments and a quasi-classical subsystem of conduction
electrons. A local magnetic exchange couples these sub-systems. Localized
magnetic moments in transition-metal chalcogenides and halides form a hexagonal
lattice. Here we focus on the N\'eel type anti-ferromagnet that is illustrated
in Fig. \ref{fig:afm_phases}a. In this case, one can define two sub-lattices A
and B that host local magnetic moments $\bs{S}^\textrm{A}$ and
$\bs{S}^\textrm{B}$, respectively. For the discussion of Gilbert damping, we
ignore the weak dependence of both fields on atomic positions and assume that
the modulus $S=|\bs{S}^{\textrm{A}(\textrm{B})}|$ is time-independent. 

Under these assumptions, the magnetization dynamics of localized moments may be
described in terms of two fields 
\be
\bs{m} = \frac{1}{2S}\big(\bs{S}^\textrm{A}+\bs{S}^\textrm{B}\big),\quad \bs{n}
= \frac{1}{2S}\big(\bs{S}^\textrm{A}-\bs{S}^\textrm{B}\big),
\e
which are referred to as the magnetization and staggered magnetization (or
N\'eel vector), respectively. Within the mean-field approach, the vector fields
yield the equations of motion
\beml
\label{eq:EOM}
\begin{align}
\label{eq:EOMa}
\dot{\bs{n}} =\,& -J\,\bs{n}\times\bs{m} + \bs{n}\times\delta\bs{s}^+ + \bs{m}\times\delta\bs{s}^-,\\
\label{eq:EOMb}
\dot{\bs{m}} =\,& \bs{m}\times\delta\bs{s}^+ + \bs{n}\times\delta\bs{s}^-,
\end{align}
\eml
where dot stands for the time derivative, while $\delta\bs{s}^+$ and
$\delta\bs{s}^-$ stand for the mean staggered and non-staggered non-equilibrium
fields that are proportional to the variation of the corresponding
spin-densities of conduction electrons caused by the time dynamics of $\bs{n}$
and $\bs{m}$ fields. The energy $J$ is proportional to the anti-ferromagnet
exchange energy for localized momenta. 

In Eqs.~(\ref{eq:EOM}) we have omitted terms that are proportional to easy axis
anisotropy for the sake of compactness. These terms are, however, important and
will be introduced later in the text. 

In the framework of Eqs.~(\ref{eq:EOM}) the Gilbert damping can be computed as
the linear response of the electron spin-density variation to a time change in
both the magnetization and the N\'eel vector (see e.\,g. Refs.
\cite{brataas_scattering_2008, ebert_ab_2011, baglai_giant_2020}).

In this definition, Gilbert damping describes the relaxation of localized spins
by transferring both total and staggered angular momenta to the lattice by means
of conduction electron scattering off impurities. Such a transfer is facilitated
by spin-orbit interaction. 
  
The structure of the full Gilbert damping tensor can be rather complicated as
discussed in Ref.~\onlinecite{baglai_giant_2020}. However, by taking into account easy
axis or easy plane anisotropy we may reduce the complexity of relevant spin
configurations to parameterize
\beml
\label{gilbert_alphas}
\begin{align}
\delta\bs{s}^+ &= \alpha_{m}^\parallel\dot{\bs{m}_\parallel} + \alpha_{m}^{\perp}\dot{\bs{m}_\perp} + \alpha_m\bs{n}_\parallel\times(\bs{n}_\parallel\times\dot{\bs{m}_\parallel}),\\
\delta\bs{s}^- &= \alpha_{n}^\parallel\dot{\bs{n}_\parallel} + \alpha_{n}^{\perp}\dot{\bs{n}_\perp}+\alpha_n\bs{n}_\parallel\times(\bs{n}_\parallel\times\dot{\bs{n}_\parallel}),
\label{eq:gilbert_staggered}
\end{align}
\eml
where the superscripts $\parallel$ and $\perp$ refer to the in-plane and
perpendicular-to-the-plane projections of the corresponding vectors,
respectively. The six coefficients $\alpha_{m}^{\parallel}$,
$\alpha_{m}^{\perp}$, $\alpha_m$,  $\alpha_{n}^{\parallel}$,
$\alpha_{n}^{\perp}$, and $\alpha_n$ parameterize the Gilbert damping.  

Inserting Eqs.~(\ref{gilbert_alphas}) into the equations of motion of
Eqs.~(\ref{eq:EOM}) produces familiar Gilbert damping terms. The damping
proportional to time-derivatives of the N\'eel vector $\bs{n}$ is in general
many orders of magnitude smaller than that proportional to the time-derivatives
of the magnetization vector $\bs{m}$
\cite{liu_mode-dependent_2017,baglai_giant_2020}. Due to the same reason, the
higher harmonics term
$\alpha_m\bs{n}_\parallel\times(\bs{n}_\parallel\times\partial_t\bs{m}_\parallel)$
can often be neglected.  

Thus, in the discussion below we may focus mostly on the coefficients
$\alpha_{m}^{\parallel}$ and $\alpha_{m}^{\perp}$ that play the most important
role in the magnetization dynamics of our system. The terms proportional to the
time-derivative of $\bs{n}$ correspond to the transfer of angular momentum
between the sub-lattices and are usually less relevant. We refer to the results
of Ref.~\onlinecite{baglai_giant_2020} when discussing these terms. 

All Gilbert damping coefficients are intimately related to the electron spin
relaxation time. The latter is relatively well understood in non-magnetic
semiconductors with spin-orbital coupling. When a conducting electron moves in a
steep potential it feels an effective magnetic field caused by relativistic
effects. Thus, in a disordered system, the electron spin is subject to a random
magnetic field each time it scatters off an impurity. At the same time, an
electron also experiences precession around an effective spin-orbit field when
it moves in between the collisions. Changes in spin direction \emph{between}
collisions are referred to as Dyakonov-Perel relaxation
\cite{dyakonov1972spin,DYAKONOV1986}, while changes in spin-direction
\emph{during} collisions are referred to as Elliot-Yafet relaxation
\cite{elliott_theory_1954,yafet_g_1963}. 

The spin-orbit field in semiconductors induces a characteristic frequency of
spin precession $\Omega_\textrm{s}$, while scalar disorder leads to a finite
transport time $\tau$ of the conducting electrons. One may, then, distinguish
two limits: (i) $\Omega_\textrm{s} \tau \ll 1$ in which case the electron does
not have sufficient time to change its direction between consecutive scattering
events (Elliot-Yafet relaxation), and (ii) $\Omega_\textrm{s}\tau \gg 1$ in
which case the electron spin has multiple precession cycles in between the
collisions (Dyakonov-Perel relaxation). 

The corresponding processes define the so-called spin relaxation time, $\tau_s$.
In a 2D system the spin life-time $\tau_s^\para$, for the in-plane spin
components, appears to be double the size of the life-time of the spin component
that is perpendicular to the plane, $\tau_s^\perp$ \cite{DYAKONOV1986}. This
geometric effect has largely been overlooked. For non-magnetic 2D semiconductor
one can estimate \cite{dyakonov_spintronics_2004, dyakonov_spin_2017}
\be
\frac{1}{\tau_s^\para} \sim 
\bc
\Omega^2_s\tau, & \Omega_\textrm{s}\tau \ll 1\\
1/\tau,  & \Omega_\textrm{s}\tau \gg 1 \ec,
\qquad
\tau_s^\para = 2 \tau_s^\perp.
\label{eq:dyakonov}
\e

A pedagogical derivation and discussion of Eq.~\ref{eq:dyakonov} can be found in
Refs.~\onlinecite{dyakonov_spintronics_2004, dyakonov_spin_2017}. Because electrons
are confined in two dimensions the random spin-orbit field is always directed
in-plane, which leads to a decrease in the in-plane spin-relaxation rate by a
factor of two compared to the out-of-plane spin-relaxation rate as demonstrated
first in Ref.~\onlinecite{DYAKONOV1986} (see Refs.~\onlinecite{averkiev_spin_2002,
burkov_theory_2004, burkov_spin_2004, sinitsyn_theory_2016, dyakonov_spin_2017}
as well). The reason is that the perpendicular-to-the-plane component of spin is
influenced by two components of the randomly changing magnetic field, i.\,e. $x$
and $y$, whereas the parallel-to-the-plane spin components are only influenced
by a single component of the fluctuating fields, i.\,e. the $x$ spin projection
is influenced only by the $y$ component of the field and vice-versa. The
argument has been further generalized in Ref.~\onlinecite{baglai_giant_2020} to the
case of strongly separated spin-orbit split Fermi surfaces. In this limit, the
perpendicular-to-the-plane spin-flip processes on scalar disorder potential
become fully suppressed. As a result, the perpendicular-to-the-plane spin
component becomes nearly conserved, which results in a giant anisotropy of
Gilbert damping in this regime. 

In magnetic systems that are, at the same time, conducting there appears to be
at least one additional energy scale, $\Delta_\textrm{sd}$, that characterizes
exchange coupling of conduction electron spin to the average magnetic moment of
localized electrons. (In the case of s-d model description it is the magnetic
exchange between the spin of conduction $s$ electron and the localized magnetic
moment of $d$ or $f$ electron on an atom.) This additional energy scale
complicates the simple picture of Eq.~(\ref{eq:dyakonov}) especially in the case
of an anti-ferromagnet. The electron spin precession is now defined not only by
spin-orbit field but also by $\Delta_\textrm{sd}$. As the result the conditions
$\Omega_\textrm{s} \tau \ll 1$ and $\Delta_\textrm{sd} \tau\gg 1$ may easily
coexist. This dissolves the distinction between Elliot-Yafet and Dyakonov-Perel
mechanisms of spin relaxation. One may, therefore, say that both Elliot-Yafet
and Dyakonov-Perel mechanisms may act simultaneously in a typical 2D metallic
magnet with spin-orbit coupling. The Gilbert damping computed from the
microscopic model that we formulate below will always contain both contributions
to spin-relaxation.

\section{Microscopic model and results}

The microscopic model that we employ to calculate Gilbert damping is the
so-called $s$--$d$ model that couples localized magnetic momenta
$\bs{S}^\textrm{A}$ and $\bs{S}^\textrm{B}$ and conducting electron spins via
the local magnetic exchange $\Delta_\textrm{sd}$. Our effective low-energy
Hamiltonian for conduction electrons reads
\be
H = v_f\, \bs{p}\cdot \bb{\Sigma} +
\frac{\lambda}{2}\,\big[\bb{\sigma}\times\bb{\Sigma}\big]_z-\Delta_\textrm{sd}\,\bs{n}\cdot\bb{\sigma}\,\Sigma_z\Lambda_z+V(\bs{r}),
\label{eq:linear_response}
\e
where the vectors $\bb{\Sigma}$, $\bb{\sigma}$ and $\bb{\Lambda}$ denote the
vectors of Pauli matrices acting on sub-lattice, spin and valley space,
respectively. We also introduce the Fermi velocity $v_f$, Rashba-type spin-orbit
interaction $\lambda$, and a random impurity potential $V(\bs{r})$. 

The Hamiltonian of Eq.~(\ref{eq:linear_response}) can be viewed as the graphene
electronic model where conduction electrons have 2D Rashba spin-orbit coupling
and are also coupled to anti-ferromagnetically ordered classical spins on the
honeycomb lattice. 

The coefficients $\alpha_{m}^{\parallel}$ and $\alpha_{m}^{\perp}$ are obtained
using linear response theory for the response of spin-density $\delta\bs{s}^+$
to the time-derivative of magnetization vector $\partial_t\bs{m}$. Impurity
potential $V(\bs{r})$ is important for describing momentum relaxation to the
lattice. This is related to the angular momentum relaxation due to spin-orbit
coupling. The effect of random impurity potential is treated perturbatively in
the (diffusive) ladder approximation that involves a summation over diffusion
ladder diagrams. The details of the microscopic calculation can be found in the
Appendices. 

Before presenting the disorder-averaged quantities
$\alpha_{m}^{\parallel,\perp}$, it is instructive to consider first the
contribution to Gilbert damping originating from a small number of
electron-impurity collisions. This clarifies how the number of impurity
scattering effects will affect the final result. 

Let us annotate the Gilbert damping coefficients with an additional superscript
$(l)$ that denotes the number of scattering events that are taken into account.
This means, in the diagrammatic language, that the corresponding quantity is
obtained by summing up the ladder diagrams with $\leq l$ disorder lines. Each
disorder line corresponds to a quasi-classical scattering event from a single
impurity. The corresponding Gilbert damping coefficient is, therefore, obtained
in the approximation where conduction electrons have scattered at most $l$
number of times before releasing their non-equilibrium magnetic moment into a
lattice.

\begin{figure}
\centerline{
\includegraphics[width=\columnwidth]{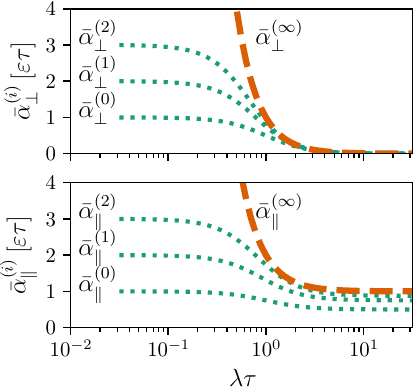}
} \caption{Gilbert damping in the limit $\Delta_\textrm{sd}=0$. Dotted (green)
lines correspond to the results of the numerical evaluation of
$\bar{\alpha}_{m,\perp,\para}^{(l)}$ for $l=0,1,2$ as a function of the
parameter $\lambda \tau$. The dashed (orange) line corresponds to the diffusive
(fully vertex corrected) results for $\bar{\alpha}_{m}^{\perp,\para.}$.}
\vspace{-0.5cm}
\label{fig:alpha_plot}
\end{figure}

To make final expressions compact we define the dimensionless Gilbert damping
coefficients $\bar{\alpha}_{m}^{\para,\perp}$ by extracting the scaling factor
\be
\label{GDD}
\alpha_{m}^{\para,\perp} =
\frac{\mathcal{A}\Delta_\textrm{sd}^2}{\pi\hbar^2v_f^2S}\;\bar{\alpha}_{m}^{\para,\perp},  
\e
where $\mathcal{A}$ is the area of the unit cell, $v_f$ is the Fermi velocity of
the conducting electrons and $\hbar=h/2\pi$ is the Planck's constant. We also
express the momentum scattering time $\tau$ in inverse energy units, $\tau\to
\hbar \tau$.  

Let us start by computing the coefficients
$\bar{\alpha}_{m}^{\parallel,\perp(l)}$ in the formal limit
$\Delta_\textrm{sd}\to 0$. We can start with the ``bare bubble'' contribution
which describes spin relaxation without a single scattering event. The
corresponding results read
\beml
\label{zeroD}
\begin{align}
\bar{\alpha}_{m,\perp }^{(0)} = \, & \ep\tau \frac{1-\lambda^2/4\ep^2}{1+\lambda^2\tau^2},\\
\bar{\alpha}_{m,\para}^{(0)} = \, &  \ep\tau\lt(\frac{1+\lambda^2\tau^2/2}{1+\lambda^2\tau^2}-\frac{\lambda^2}{8\ep^2}\rt),
\end{align}
\eml
where $\ep$ denotes the Fermi energy which we consider positive (electron-doped
system). 

In all realistic cases, we have to consider $\lambda/\ep \ll 1$, while the
parameter $\lambda \tau$ may in principle be arbitrary. For $\lambda \tau\ll 1$
the disorder-induced broadening of the electron Fermi surfaces exceeds the
spin-orbit induced splitting. In this case one basically finds no anisotropy of
``bare'' damping: $\bar{\alpha}_{m,\perp }^{(0)} =
\bar{\alpha}_{m,\para}^{(0)}$. In the opposite limit of substantial spin-orbit
splitting one gets an ultimately anisotropic damping $\bar{\alpha}_{m,\perp
}^{(0)} \ll \bar{\alpha}_{m,\para}^{(0)}$. This asymptotic behavior can be
summarized as
\beml
\label{ZeroSC}
\begin{align}
\bar{\alpha}_{m,\perp }^{(0)}  & = \ep\tau
\bc 1  & \lambda\tau\ll1,\\
 (\lambda\tau)^{-2} & \lambda\tau \gg 1,
\ec\\
\bar{\alpha}_{m,\para}^{(0)}  & = \ep\tau 
\bc 1  &  \lambda\tau\ll1,\\
\frac{1}{2} \lt(1 + (\lambda\tau)^{-2}\rt) & \lambda\tau \gg 1,
\ec
\end{align}
\eml
where we have used that $\ep\gg \lambda$. 

The results of Eq.~(\ref{ZeroSC}) modify by electron diffusion. By taking into
account up to $l$ scattering events we obtain
\beml
\label{Levents}
\begin{align}
\label{eq:perpi}
\bar{\alpha}_{m,\perp }^{(l)}  & = \ep\tau 
\bc
l + \mathcal{O}(\lambda^2\tau^2)& \lambda \tau \ll 1,\\
 (1+\delta_{l0})/(\lambda\tau)^{2} & \lambda \tau\gg 1,
\ec \\
\label{eq:parai}
\bar{\alpha}_{m,\para}^{(l)}  &  =\ep\tau 
\bc
l + \mathcal{O}(\lambda^2\tau^2)& \lambda \tau\ll 1,\\
1-(1/2)^{l+1}+\mathcal{O}(\lt(\lambda\tau\rt)^{-2}) & \lambda\tau\gg 1,
\ec
\end{align}
\eml
where we have used $\ep\gg \lambda$ again. 

From Eqs.~(\ref{Levents}) we see that the Gilbert damping for $\lambda\tau\ll 1$
gets an additional contribution of $\ep\tau$ from each scattering event as
illustrated numerically in Fig.~\ref{fig:alpha_plot}. This leads to a formal
divergence of Gilbert damping in the limit  $\lambda \tau \ll 1$. While, at
first glance, the divergence looks like a strong sensitivity of damping to
impurity scattering, in reality, it simply reflects a diverging spin life-time.
Once a non-equilibrium magnetization $\bs{m}$ is created it becomes almost
impossible to relax it to the lattice in the limit of weak spin-orbit coupling.
The formal divergence of $\alpha_{m}^{\perp}=\alpha_{m}^{\para}$ simply reflects
the conservation law for electron spin polarization in the absence of spin-orbit
coupling such that the corresponding spin life-time becomes arbitrarily large as
compared to the momentum scattering time $\tau$. 

By taking the limit $l\to\infty$ (i.\,e. by summing up the entire diffusion
ladder) we obtain compact expressions 
\beml
\label{zeroDelta}
\begin{align}
\label{eq:alphaperpzerodelta}
\bar{\alpha}_{m}^{\perp}\equiv\bar{\alpha}_{m,\perp }^{(\infty)}  & = \ep \tau\, \frac{1}{2\lambda^2\tau^2},\\
\label{eq:alphaparallelzerodelta}  
\bar{\alpha}_{m}^{\para}\equiv\bar{\alpha}_{m,\para}^{(\infty)}  & = \ep \tau\,\frac{1+\lambda^2\tau^2}{\lambda^2\tau^2},
\end{align}
\eml
which assume $\bar{\alpha}_{m}^{\perp}\ll \bar{\alpha}_{m}^{\para}$ for
$\lambda\tau \gg 1$ and $\bar{\alpha}_{m}^{\perp} = \bar{\alpha}_{m}^{\para}/2$
for $\lambda\tau \ll 1$. The factor of $2$ difference that we observe when
$\lambda\tau \ll 1$, corresponds to a difference in the electron spin life-times
$\tau_s^\perp=\tau_s^\para/2$ that was discussed in the introduction
\cite{DYAKONOV1986}. 

Strong spin-orbit coupling causes a strong out-of-plane anisotropy of damping,
$\bar{\alpha}_{m}^{\perp}\ll \bar{\alpha}_{m}^{\para}$ which corresponds to a
suppression of the perpendicular-to-the-plane damping component. As a result,
the spin-orbit interaction makes it much easier to relax the magnitude of the
$m_z$ component of magnetization than that of in-plane components. 

\begin{figure}
\centerline{
\includegraphics[width=\columnwidth]{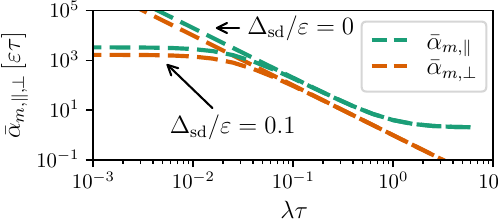}
} \caption{Numerical results for the Gilbert damping components in the diffusive
limit (vertex corrected)as the function of the spin-orbit coupling strength
$\lambda$. The results correspond to $\ep\tau=50$ and
$\Delta_\textrm{sd}\tau=0.1$ and agree with the asymptotic expressions of
Eq.~(\ref{finiteD}). Three different regimes can be distinguished for
$\bar{\alpha}_{m}^\para$: i) spin-orbit independent damping
$\bar{\alpha}_{m}^\para \propto \ep^3\tau/\Delta_\textrm{sd}^2$ for the exchange
dominated regime, $\lambda\tau \ll \Delta_\textrm{sd}/\ep$, ii) the damping
$\bar{\alpha}_{m}^\para \propto \ep/\lambda^2\tau$ for Elliot-Yafet relaxation
regime, $\Delta_\textrm{sd}/\ep\ll \lambda\tau \ll 1$, and  iii) the damping
$\bar{\alpha}_{m}^\para \propto \ep\tau$ for the Dyakonov-Perel relaxation
regime, $ \lambda\tau\gg 1$. The latter regime is manifestly absent for
$\bar{\alpha}_{m}^\perp$ in accordance with Eqs.~(\ref{SRpara},\ref{SRperp}). }
\label{fig:alpha_plot_full}
\end{figure}

Let us now turn to the dependence of $\bar{\alpha}_{m}$ coefficients on
$\Delta_\textrm{sd}$ that is illustrated numerically in
Fig.~\ref{fig:alpha_plot_full}. We consider first the case of absent spin-orbit
coupling $\lambda=0$. In this case, the combination of spin-rotational and
sub-lattice symmetry (the equivalence of A and B sub-lattice) must make Gilbert
damping isotropic (see e.\,g. \cite{Kamra2018,baglai_giant_2020}). The direct
calculation for $\lambda=0$ does, indeed, give rise to the isotropic result
$\bar{\alpha}_{m}^{\perp}=\bar{\alpha}_{m}^{\para}=\ep\tau(\ep^2+\Delta_\textrm{sd}^2)/2\Delta_\textrm{sd}^2$,
which is, however, in contradiction to the limit $\lambda\to 0$ in
Eq.~(\ref{zeroDelta}). 

\begin{figure}
\centerline{
\includegraphics[width=\columnwidth]{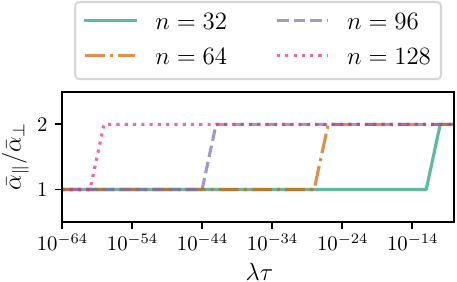}
} \caption{Numerical evaluation of Gilbert damping anisotropy in the limit
$\lambda\to 0$. Isotropic damping tensor is restored only if $\lambda=0$ with
ultimate numerical precision. The factor of $2$ anisotropy emerges at any finite
$\lambda$, no matter how small it is, and only depends on the numerical
precision $n$, i.e. the number of digits contained in each variable during
computation. The crossover from isotropic to anisotropic damping can be
understood only by considering finite, though vanishingly small, magnon $q$
vectors.}
\label{fig:num_test}
\end{figure}

At first glance, this contradiction suggests the existence of a certain energy
scale for $\lambda$ over which the anisotropy emerges. The numerical analysis
illustrated in Fig.~\ref{fig:num_test} reveals that this scale does not depend
on the values of $1/\tau$, $\Delta_\textrm{sd}$, or $\ep$. Instead, it is
defined solely by numerical precision. In other words, an isotropic Gilbert
damping is obtained only when the spin-orbit strength $\lambda$ is set below the
numerical precision in our model. We should, therefore, conclude that the
transition from isotropic to anisotropic (factor of 2) damping occurs exactly at
$\lambda=0$. Interestingly, the factor of $2$ anisotropy is absent in
Eqs.~(\ref{ZeroSC}) and emerges only in the diffusive limit. 

We will see below that this paradox can only be resolved by analyzing the
Gilbert damping beyond the infinite wave-length limit. 

One can see from Fig.~\ref{fig:alpha_plot_full} that the main effect of finite
$\Delta_\textrm{sd}$ is the regularization of the Gilbert damping divergency
$(\lambda\tau)^{-2}$ in the limit $\lambda \tau \ll 1$. Indeed, the limit of
weak spin-orbit coupling is non-perturbative for $\Delta_\textrm{sd}/\ep\ll
\lambda\tau \ll 1$, while, in the opposite limit, $\lambda\tau \ll
\Delta_\textrm{sd}/\ep\ll1$, the results of Eqs.~(\ref{zeroDelta}) are no longer
valid. Assuming  $\Delta_\textrm{sd}/\ep\ll 1$ we obtain the asymptotic
expressions for the results presented in Fig.~\ref{fig:alpha_plot_full} as
\beml
\label{finiteD}
\begin{align}
\label{eq:alphaperp}
\bar{\alpha}_{m}^{\perp} &=\frac{1}{2}\ep\tau\,
\bc        
\frac{2}{3}\frac{\ep^2+\Delta_\textrm{sd}^2}{\Delta_\textrm{sd}^2}& \lambda\tau \ll \Delta_\textrm{sd}/\ep,\\[4pt]
\frac{1}{\lambda^2\tau^2}& \lambda\tau \gg \Delta_\textrm{sd}/\ep, 
\ec\\
\label{eq:alphapara}
\bar{\alpha}_{m}^{\para} &=\ep\tau\, 
\bc    
\frac{2}{3}\frac{\ep^2+\Delta_\textrm{sd}^2}{\Delta_\textrm{sd}^2} & \lambda\tau \ll \Delta_\textrm{sd}/\ep, \\[4pt]
1+\frac{1}{\lambda^2\tau^2}& \lambda\tau \gg \Delta_\textrm{sd}/\ep,
\ec
\end{align}
\eml
which suggest that $\bar{\alpha}_{m}^{\perp}/\bar{\alpha}_{m}^{\para}=2$ for
$\lambda\tau\ll 1$. In the opposite limit, $\lambda\tau\gg 1$, the anisotropy of
Gilbert damping grows as
$\bar{\alpha}_{m}^{\para}/\bar{\alpha}_{m}^{\perp}=2\lambda^2\tau^2$. 

The results of Eqs.~(\ref{finiteD}) can also be discussed in terms of the
electron spin life-time,
$\tau_s^{\perp(\para)}=\bar{\alpha}_{m}^{\perp(\para)}/\ep$. For the inverse
in-plane spin life-time we find
\be
\label{SRpara}
\frac{1}{\tau_s^\para}=\bc 3\Delta_\textrm{sd}^2/2\ep^2\tau & \lambda\tau \ll
\Delta_\textrm{sd}/\ep,\\
\lambda^2\tau & \Delta_\textrm{sd}/\ep \ll  \lambda\tau \ll 1,\\
1/\tau & 1\ll \lambda\tau,  
\ec
\e
that, for $\Delta_\textrm{sd}=0$, is equivalent to the known result of
Eq.~(\ref{eq:dyakonov}). Indeed, for $\Delta_\textrm{sd} = 0$, the magnetic
exchange plays no role and one observes the cross-over from Elliot-Yafet
($\lambda\tau \ll 1$) to Dyakonov-Perel ($\lambda\tau \gg 1$) spin relaxation. 

This cross-over is, however, absent in the relaxation of the perpendicular spin
component
\be
\label{SRperp}
\frac{1}{\tau_s^\perp}= 2 \bc 3\Delta_\textrm{sd}^2/2\ep^2\tau & \lambda\tau \ll
\Delta_\textrm{sd}/\ep,\\
\lambda^2\tau & \Delta_\textrm{sd}/\ep \ll  \lambda\tau,  
\ec
\e
where Elliot-Yafet-like relaxation extends to the regime $\lambda\tau \gg 1$. 

As mentioned above, the factor of two anisotropy in spin-relaxation of $2D$
systems, $\tau^\para_s = 2\tau_s^\perp$, is known in the literature
\cite{DYAKONOV1986} (see Refs. \cite{averkiev_spin_2002, burkov_theory_2004,
dyakonov_spin_2017} as well). Unlimited growth of spin life-time anisotropy,
$\tau_s^\para/\tau_s^\perp=2\lambda^2\tau^2$, in the regime $\lambda\tau\ll 1$
has been described first in Ref.~\onlinecite{baglai_giant_2020}. It can be
qualitatively explained by a strong suppression of spin-flip processes for $z$
spin component due to spin-orbit induced splitting of Fermi surfaces. The
mechanism is effective only for scalar (non-magnetic) disorder. Even though such
a mechanism is general for any magnetic or non-magnetic 2D material with
Rashba-type spin-orbit coupling, the effect of the spin life-time anisotropy on
Gilbert damping is much more relevant for anti-ferromagnets. Indeed, in an
anti-ferromagnetic system the modulus of $\bs{m}$ is, by no means, conserved,
hence the variations of perpendicular and parallel components of the
magnetization vector are no longer related. 

In the regime, $\lambda\tau \ll \Delta_\textrm{sd}/\ep$ the spin life-time is
defined by exchange interaction and the distinction between Dyakonov-Perel and
Elliot-Yafet mechanisms of spin relaxation is no longer relevant. In this
regime, the spin-relaxation time is by a factor $(\ep/\Delta_\textrm{sd})^2$
larger than the momentum relaxation time. 

Let us now return to the problem of emergency of the factor of $2$ anisotropy of
Gilbert damping at $\lambda=0$. We have seen above (see Fig.~\ref{fig:num_test})
that, surprisingly, there exists no energy scale for the anisotropy to emerge.
The transition from the isotropic limit ($\lambda=0$) to a finite anisotropy
appeared to take place exactly at $\lambda=0$. We can, however, generalize the
concept of Gilbert damping by considering the spin density response function at
a finite wave vector $\bs{q}$. 

To generalize the Gilbert damping, we are seeking a response of spin density at
a point $\bs{r}$, $\delta\bs{s}_+(\bs{r})$ to a time derivative of magnetization
vectors $\dot{\bs{m}}_\para$ and $\dot{\bs{m}}_\perp$ at the point $\bs{r}'$.
The Fourier transform with respect to $\bs{r}-\bs{r}'$ gives the Gilbert damping
for a magnon with the wave-vector $\bs{q}$.   

The generalization to a finite $\bs{q}$-vector shows that the limits $\lambda\to
0$ and $q\to 0$ cannot be interchanged. When the limit $\lambda\to 0$ is taken
before the limit $q\to 0$ one finds an isotropic Gilbert damping, while for the
opposite order of limits, it becomes a factor of $2$ anisotropic. In a realistic
situation, the value of $q$ is limited from below by an inverse size of a
typical magnetic domain $1/L_\textrm{m}$, while the spin-orbit coupling is
effective on the length scale $L_\lambda = 2\pi \hbar v_f /\lambda$. In this
picture, the isotropic Gilbert damping is characteristic for the case of
sufficiently small domain size $L_\textrm{m} \ll L_\lambda$, while the
anisotropic Gilbert damping corresponds to the case $L_\lambda \ll L_m$. 

In the limit $q\ell\ll 1$, where $\ell=v_f\tau$ is the electron mean free path,
we can summarize our results as
\beml
\label{finiteQ}
\begin{align}
\label{eq:alphaperp2}
\bar{\alpha}_{m}^{\perp} &= \ep\tau\,
\bc
\frac{\ep^2+\Delta_\textrm{sd}^2}{2\Delta_\textrm{sd}^2} & \lambda\tau \ll q\ell \ll \Delta_\textrm{sd}/\ep, \\
\frac{1}{3}\frac{\ep^2+\Delta_\textrm{sd}^2}{\Delta_\textrm{sd}^2} & q\ell \ll \lambda\tau \ll \Delta_\textrm{sd}/\ep,\\
\frac{1}{2\lambda^2\tau^2} & \lambda\tau \gg \Delta_\textrm{sd}/\ep, 
\ec,\\
\label{eq:alphapara2}
\bar{\alpha}_{m}^{\para} &= 
\ep\tau\,
\bc
\frac{\ep^2+\Delta_\textrm{sd}^2}{2\Delta_\textrm{sd}^2} & \lambda\tau \ll q\ell \ll \Delta_\textrm{sd}/\ep, \\
\frac{2}{3}\frac{\ep^2+\Delta_\textrm{sd}^2}{\Delta_\textrm{sd}^2} & q\ell \ll \lambda\tau \ll \Delta_\textrm{sd}/\ep,\\
1+\frac{1}{\lambda^2\tau^2} & \lambda\tau \gg \Delta_\textrm{sd}/\ep, 
\ec
\end{align}
\eml
which represent a simple generalization of Eqs.~(\ref{finiteD}).

\begin{figure}
\centerline{\includegraphics[width=\columnwidth]{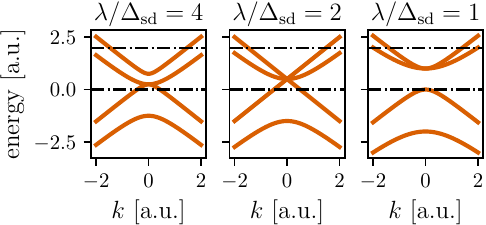}}
\caption{Band-structure for the effective model of
Eq.~(\ref{eq:linear_response}) in a vicinity of $\bs{K}$ valley assuming
$n_z=1$. Electron bands touch for $\lambda=2\Delta_\textrm{sd}$. The regime
$\lambda\leq2\Delta_\textrm{sd}$ corresponds to spin-orbit band inversion. The
band structure in the valley $\bs{K}'$ is inverted. Our microscopic analysis is
performed in the electron-doped regime for the Fermi energy above the gap as
illustrated by the top dashed line. The bottom dashed line denotes zero energy
(half-filling).}
\label{fig:bands}
\end{figure}

The results of Eqs.~(\ref{finiteQ}) correspond to a simple behavior of Gilbert
damping anisotropy,
\be
\label{eq:eta1}
\bar{\alpha}_{m}^{\para}/\bar{\alpha}_{m}^{\perp} = \bc 1& \lambda\tau\ll
q\ell,\\
2\lt(1+\lambda^2\tau^2\rt)&  q\ell\ll \lambda\tau,
\ec
\e
where we still assume $q\ell \ll 1$. 

\section{Anti-ferromagnetic resonance}

The broadening of the anti-ferromagnet resonance peak is one obvious quantity that
is sensitive to Gilbert damping. The broadening is however not solely defined by
a particular Gilbert damping component but depends also on both magnetic
anisotropy and anti-ferromagnetic exchange. 

To be more consistent we can use the model of Eq.~(\ref{eq:linear_response}) to
analyze the contribution of conduction electrons to an easy axis anisotropy. The
latter is obtained by expanding the free energy for electrons in the value of
$n_z$, which has a form $E = -Kn_z^2/2$. With the conditions $\ep/\lambda\gg 1$
and $\ep/\Delta_\textrm{sd}\gg 1$ we obtain the anisotropy constant as
\be
\label{Kvalue}
K= \frac{\mathcal{A}}{2\pi\hbar^2v^2}
\bc
\Delta_\textrm{sd}^2\lambda  & 2\Delta_\textrm{sd}/\lambda \leq 1, \\
\Delta_\textrm{sd}\lambda^2/2  & 2\Delta_\textrm{sd}/\lambda \geq 1,
\ec
\e
where $\mathcal{A}$ is the area of the unit cell. Here we assume both $\lambda$
and $\Delta_\textrm{sd}$ positive, therefore, the model naturally gives rise to
an easy axis anisotropy with $K>0$. In real materials, there exist other sources
of easy axis or easy plane anisotropy. In-plane magneto-crystalline anisotropy
also plays an important role.  For example, N\'eel-type anti-ferromagnets with
easy-axis anisotropy are FePS$_3$, FePSe$_3$ or MnPS$_3$, whereas those with
easy plane and in-plane magneto-crystalline anisotropy are NiPS$_3$ and
MnPSe$_3$. Many of those materials are, however, Mott insulators. Our
qualitative theory may still apply to materials like MnPS$_3$ monolayers at
strong electron doping.   

The transition from $2\Delta_\textrm{sd}/\lambda \geq 1$ to
$2\Delta_\textrm{sd}/\lambda \leq 1$ in Eq.~(\ref{Kvalue}) corresponds to the
touching of two bands in the model of Eq.~(\ref{eq:linear_response}) as
illustrated in Fig.~\ref{fig:bands}. 

Anti-ferromagnetic magnon frequency and life-time in the limit $q\to 0$ are
readily obtained by linearizing the equations of motion
\beml
\label{AFMEOM2}
\begin{align}
\label{ndot2}
\dot{\bs{n}} = & \, -J\, \bs{n}\!\times\!\bs{m} +K\, \bs{m}\!\times\!\bs{n}_\perp
+\bs{n}\!\times\!\lt(\hat{\alpha}_m\dot{\bs{m}}\rt),\\
\label{mdot2}
\dot{\bs{m}} = & \,K\,\bs{n}\!\times\!\bs{n}_\perp+\bs{n} \times \lt( \hat{\alpha}_n\dot{\bs{n}}\rt),
\end{align}
\eml
where we took into account easy axis anisotropy $K$ and disregarded irrelevant
terms $\bs{m}\!\times\! \lt(\hat{\alpha}_n\dot{\bs{n}}\rt)$ and $\bs{m} \times
\lt( \hat{\alpha}_m\dot{\bs{m}}\rt)$. We have also defined Gilbert damping
tensors such as $\hat{\alpha}_m\dot{\bs{m}} =\alpha^\para_m\dot{\bs{m}}_\para
+\alpha^\perp_m\dot{\bs{m}}_\perp$, $\hat{\alpha}_n\dot{\bs{n}}
=\alpha^\para_n\dot{\bs{n}}_\para +\alpha^\perp_n\dot{\bs{n}}_\perp$.

In the case of easy axis anisotropy we can use the linearized modes
$\bs{n}=\hat{\bs{z}}+\delta\bs{n}_\para\,e^{i\omega t}$,
$\bs{m}=\delta\bs{m}_\para\,e^{i\omega t}$, hence we get the energy of $q=0$
magnon as 
\begin{align}
\label{omega}
&\omega = \omega_0-i\Gamma/2,\\
\label{omega0}
&\omega_0=\sqrt{JK},\qquad \Gamma=J\alpha^\para_n+K \alpha^\para_m
\end{align}
where we took into account that $K\ll J$. The expression for $\omega_0$ is well
known due to Kittel and Keffer \cite{PhysRev.82.565, PhysRev.85.329}.  

Using Ref.~\onlinecite{baglai_giant_2020} we find out that $\alpha_n^\para \simeq
\alpha_m^\perp (\lambda/\ep)^2$ and $\alpha_n^\perp \simeq \alpha_m^\para
(\lambda/\ep)^2$, hence
\be
\label{G}
\Gamma \simeq \alpha_m^\para\lt(K + \frac{J/2}{\ep^2/\lambda^2+\ep^2\tau^2}\rt),
\e
where we have simply used Eqs.~\eqref{zeroDelta}.  Thus, one may often ignore
the contribution $J\alpha^\para_n$ as compared to $K \alpha^\para_m$ despite the
fact that $K\ll J$. 

In the context of anti-ferromagnets, spin-pumping terms are usually associated
with the coefficients $\alpha^\para_n$ in Eq.~\eqref{eq:gilbert_staggered} that
are not in the focus of the present study. Those coefficients have been analyzed
for example in Ref. \onlinecite{baglai_giant_2020}. In this manuscript we simply use
the known results for $\alpha_n$ in Eqs.~(\ref{AFMEOM2}-\ref{omega0}), where we
illustrate the effect of both spin-pumping coefficient $\alpha_n$ and the direct
Gilbert damping $\alpha_m$ on the magnon life time. One can see from
Eqs.~(\ref{omega0},\ref{G}) that the spin-pumping contributions do also
contribute, though indirectly, to the magnon decay. The spin pumping
contributions become more important in magnetic materials with small magnetic
anisotropy.  The processes characterized by the coefficients $\alpha_n$ may also
be interpreted in terms of angular momentum transfer from one AFM sub-lattice to
another. In that respect, the spin pumping is specific to AFM, and is
qualitatively different from the direct Gilbert damping processes ($\alpha_m$)
that describe the direct momentum relaxation to the lattice. 

\begin{figure}
\centerline{\includegraphics[width=\columnwidth]{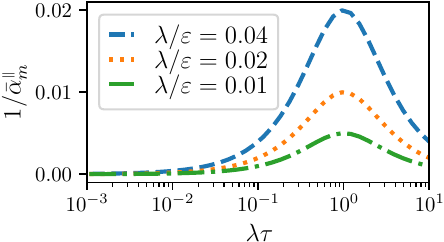}}
\caption{Numerical evaluation of the inverse Gilbert damping
$1/\bar{\alpha}_{m}^{\para}$ as a function of the momentum relaxation time
$\tau$. The inverse damping is peaked at $\tau\propto 1/\lambda$ which also
corresponds to the maximum of the anti-ferromagnetic resonance quality factor in
accordance with Eq.~(\ref{QQ}). }
\label{fig:max}
\end{figure}

As illustrated in Fig.~\ref{fig:max} the quality factor of the
anti-ferromagnetic resonance (for a metallic anti-ferromagnet with easy-axis
anisotropy) is given by
\be
\label{QQ}
Q=\frac{\omega_0}{\Gamma} \simeq \frac{1}{\alpha_m^\parallel}
\sqrt{\frac{J}{K}}.
\e
Interestingly, the quality factor defined by Eq.~(\ref{QQ}) is maximized for
$\lambda \tau \simeq 1$, i.\,e. for the electron spin-orbit length being of the
order of the scattering mean free path.  

The quantities $1/\sqrt{K}$ and $1/\bar{\alpha}_{m}^{\para}$ are illustrated in
Fig.~\ref{fig:max} from the numerical analysis. As one would expect, the quality
factor vanishes in both limits $\lambda\to 0$ and $\lambda\to\infty$. The former
limit corresponds to an overdamped regime hence no resonance can be observed.
The latter limit corresponds to a constant $\alpha_m^\para$, but the resonance
width $\Gamma$ grows faster with $\lambda$ than $\omega_0$ does, hence the
vanishing quality factor. 

It is straightforward to check that the results of Eqs.~(\ref{G},\ref{QQ})
remain consistent when considering systems with either easy-plane or in-plane
magneto-crystalline anisotropy. Thus, the coefficient $\alpha^\perp_m$ normally
does not enter the magnon damping, unless the system is brought into a vicinity
of spin-flop transition by a strong external field.

%
%

\section{Conclusion}

In conclusion, we have analyzed the Gilbert damping tensor in a model of a
two-dimensional anti-ferromagnet on a honeycomb lattice. We consider the damping
mechanism that is dominated by a finite electron spin life-time due to a
combination of spin-orbit coupling and impurity scattering of conduction
electrons. In the case of a 2D electron system with Rashba spin-orbit coupling
$\lambda$, the Gilbert damping tensor is characterized by two components
$\alpha_m^\para$ and $\alpha_m^\perp$. We show that the anisotropy of Gilbert
damping depends crucially on the parameter $\lambda\tau$, where $\tau$ is the
transport scattering time for conduction electrons. For $\lambda\tau \ll 1$ the
anisotropy is set by a geometric factor of $2$, $\alpha_m^\para = 2
\alpha_m^\perp$, while it becomes infinitely large in the opposite limit,
$\alpha_m^\para = (\lambda\tau)^2 \alpha_m^\perp$ for $\lambda\tau \gg 1$.
Gilbert damping becomes isotropic exactly for $\lambda=0$, or, strictly
speaking, for the case $\lambda\ll \hbar v_f q$, where $q$ is the magnon wave
vector. 

This factor of $2$ is essentially universal, and is a geometric effect: the
z-component relaxation results from fluctuations in two in-plane spin
components, whereas in-plane relaxation stems from fluctuations of the
z-component alone. This reflects the subtleties of our microscopic model, where
the mechanism for damping is activated by the decay of conduction electron
momenta, linked to spin-relaxation through spin-orbit interactions.

We find that Gilbert damping is insensitive to magnetic order for $\lambda\gg
\Delta_\textrm{sd}/\ep\tau$, where $\Delta_\textrm{sd}$ is an effective exchange
coupling between spins of conduction and localized electrons. In this case, the
electron spin relaxation can be either dominated by scattering (Dyakonov-Perel
relaxation) or by spin-orbit precession (Elliot-Yafet relaxation). We find that
the Gilbert damping component $\alpha_m^\perp \simeq \ep/\lambda^2\tau$ is
dominated by Elliot-Yafet relaxation irrespective of the value of the parameter
$\lambda \tau$, while the other component crosses over from  $\alpha_m^\para
\simeq \ep/\lambda^2\tau$ (Elliot-Yafet relaxation) for $\lambda\tau\ll 1$, to
$\alpha_m^\para \simeq \ep\tau$ (Dyakonov-Perel relaxation) for $\lambda\tau\gg
1$. For the case  $\lambda\ll \Delta_\textrm{sd}/\ep\tau$ the spin relaxation is
dominated by interaction with the exchange field. 

Crucially, our results are not confined solely to the N\'eel order on the
honeycomb lattice: we anticipate a broader applicability across various magnetic
orders, including the zigzag order. This universality stems from our focus on
the large magnon wavelength limit. The choice of the honeycomb lattice arises
from its unique ability to maintain isotropic electronic spectra within the
plane, coupled with the ability to suppress anisotropy concerning in-plane spin
rotations. Strong anisotropic electronic spectra would naturally induce strong
anisotropic in-plane Gilbert damping, which are absent in our results.

Finally, we show that the anti-ferromagnetic resonance width is mostly defined
by $\alpha_m^\para$ and demonstrate that the resonance quality factor is
maximized for $\lambda\tau \approx 1$. Our microscopic theory predictions may be
tested for systems such as MnPS$_3$ monolayer on Pt and similar
heterostructures.

\begin{acknowledgments}

We are grateful to O.~Gomonay, R.~Duine, J.~Sinova, and A.~Mauri for helpful
discussions. This project has received funding from the European Union’s Horizon
2020 research and innovation program under the Marie Sklodowska-Curie grant
agreement No 873028. 

\end{acknowledgments}

\appendix


\section{Microscopic framework}

The microscopic model that we employ to calculate Gilbert damping belongs to a
class of so-called $s$--$d$ models that describe the physical system in the form
of a Heisenberg model for localized spins and a tight-binding model for
conduction electrons that are weakly coupled by a local magnetic exchange
interaction of the strength $\Delta_\textrm{sd}$. 

Our effective electron Hamiltonian for a metallic hexagonal anti-ferromagnet is
given by \cite{baglai_giant_2020}
\be
H_0 = v_f \bs{p}\cdot\bb{\Sigma}
+\frac{\lambda}{2}\lt[\bb{\sigma}\times\bb{\Sigma}\rt]_{z}-\Delta_\textrm{sd}\bb{n}\cdot\bb{\sigma}\,\Sigma_z\Lambda_z,
\label{sup:eq:Ham}
\e
where the vectors $\bb{\Sigma}$, $\bb{\sigma}$ and $\bb{\Lambda}$ denote the
vectors of Pauli-matrices acting on sub-lattice, spin and valley space
respectively. We also introduce the Fermi velocity $v_f$, Rashba-type spin-orbit
interaction $\lambda$. 

To describe Gilbert damping of the localized field $\bb{n}$ we have to add the
relaxation mechanism. This is provided in our model by adding a weak impurity
potential $ H=H_0+V(\bs{r})$. The momentum relaxation due to scattering on
impurities leads indirectly to the relaxation of Heisenberg spins due to the
presence of spin-orbit coupling and exchange couplings. 

For modeling the impurity potential, we adopt a delta-correlated random
potential that corresponds to the point scatter approximation, where the range
of the impurity potential is much shorter than that of the mean free path (see
e.g. section 3.8 of Ref.~\onlinecite{rammer_quantum_1986}), i.e.  
\be
\la V(\bs{r})V(\bs{r}') \ra = 2\pi \alpha (\hbar v_f)^2\delta(\bs{r}-\bs{r}'),
\e
where the dimensionless coefficient $\alpha \ll 1$ characterizes the disorder
strength. The corresponding scattering time for electrons is obtained as
$\tau=\hbar/\pi\alpha \epsilon$, which is again similar to the case of graphene. 

The response of symmetric spin-polarization $\delta \bb{s}^+$ to the
time-derivative of non-staggered magnetization, $\partial_t \bb{m}$, is defined
by the linear relation
\be
\delta s_\alpha^+ = \s_\beta \lt.\mathcal{R}_{\alpha\beta}\rt|_{\omega=0}
\dot{m}_\beta,
\e
where the response tensor is taken at zero frequency
\cite{ado_microscopic_2017,baglai_giant_2020}.  The linear response is defined
generally by the tensor 
\be
\mathcal{R}_{\alpha\beta} =\frac{\mathcal{A}\Delta^2_\textrm{sd}}{2\pi S} \int
\!\frac{\mathrm{d}\bb{p}}{(2\pi\hbar)^2} \lt\la \tr\lt[G^\textrm{R}_{\ep,
\bs{p}} \sigma_\alpha G^\textrm{A}_{\ep+\hbar\omega, \bs{p}}\sigma_\beta\rt]
\rt\ra, 
\label{sup_eq:spindensity1}
\e
where $G^\text{R(A)}_{\ep,\bb{p}}$ are standing for retarded(advanced) Green
functions and the angular brackets denote averaging over disorder fluctuations. 

The standard recipe for disorder averaging is the diffusive approximation
\cite{Rammer, mahan2013many} that is realized by replacing the bare Green
functions in Eq.~(\ref{sup_eq:spindensity1})  with disorder-averaged Green
functions and by replacing one of the vertex operators $\sigma_x$ or $\sigma_y$
with the corresponding \emph{vertex-corrected} operator that is formally
obtained by summing up ladder impurity diagrams (diffusons). 

In models with spin-orbit coupling, the controllable diffusive approximation for
non-dissipative quantities may become, however, more involved as was noted first
in Ref.~\onlinecite{ivan}. For Gilbert damping it is, however, sufficient to
consider the ladder diagram contributions only. 

The disorder-averaged Green function is obtained by including an imaginary part
of the self-energy $\Sigma^\textrm{R}$ (not to be confused here with the Pauli
matrix $\Sigma_{0,x,y,z}$) that is evaluated in the first Born approximation
\be
\im\Sigma^\textrm{R} = 2\pi\alpha v_f^2 \int\!\frac{\mathrm{d}\bb{p}}{(2\pi
)^2}\,\im \frac{1}{\ep-H_0+ i 0}.
\label{sup:eq:Sigma}
\e
The real part of the self-energy leads to the renormalization of the energy
scales $\ep$, $\lambda$ and $\Delta_\textrm{sd}$. 

In the first Born approximation, the disorder-averaged Green function is given
by
\be
G_{\ep,\bb{p}}^\text{R} = \frac{1}{\ep-H_0 - i \im \Sigma^R}.
\e

The vertex corrections are computed in the diffusive approximation. The latter
involves replacing the vertex $\sigma_\alpha$ with the \emph{vertex-corrected}
operator, 
\be
\sigma_\alpha^\textrm{vc} = \s_{l=0}^\infty \sigma_\alpha^{(l)},
\label{sup:eq:sigmavs}
\e
where the index $l$ corresponds to the number of disorder lines in the ladder. 

The operators $\sigma_\alpha^{(l)}$ can be defined recursively as  
\be
\sigma_\alpha^{(l)} = \frac{2\hbar v_f^2}{\ep
\tau}\int\!\frac{\mathrm{d}\bb{p}}{(2\pi)^2} G^\text{R}_{\ep, \bb{p}}
\sigma_\alpha^{(l-1)} G^\text{A}_{\ep+\hbar\omega, \bb{p}},
\label{sup:eq:sigmasum}
\e
where $\sigma_\alpha^{(0)} = \sigma_\alpha$.

The summation in Eq.~(\ref{sup:eq:sigmavs}) can be computed in the full operator
basis, $B_{i=\{\alpha,\beta,\gamma\}}=\sigma_\alpha\Sigma_\beta\Lambda_\gamma$,
where each index $\alpha$, $\beta$ and $\gamma$ takes on $4$ possible values
(with zero standing for the unity matrix). We may always normalize $\tr B_i B_j
= 2\delta_{ij}$ in an analogy to the Pauli matrices. The operators $B_i$ are,
then, forming a finite-dimensional space for the recursion of
Eq.~(\ref{sup:eq:sigmasum}). 

The vertex-corrected operators $B_i^\textrm{vc}$ are obtained by summing up the
matrix geometric series 
\be
B_i^\textrm{vc} = \s_j \lt(\frac{1}{1-\mathcal{F}}\rt)_{ij}B_j,
\e 
where the entities of the matrix $\mathcal{F}$ are given by
\be
\mathcal{F}_{ij} = \frac{\hbar v_f^2}{\ep
\tau}\int\!\frac{\mathrm{d}\bb{p}}{(2\pi)^2} \tr\lt[G^\text{R}_{\ep, \bb{p}} B_i
G^\text{A}_{\ep+\hbar\omega, \bb{p}} B_j\rt].
\label{sup:eq:F}
\e
Our operators of interest $\sigma_x$ and $\sigma_y$ can always be decomposed in
the operator basis as
\be
\sigma_\alpha= \frac{1}{2}\s_i B_i \,\tr\lt(\sigma_\alpha B_i\rt),
\e
hence the vertex-corrected spin operator is given by
\be
\sigma^{\textrm{vc}}_\alpha =\frac{1}{2}\s_{ij}B^{\textrm{vc}}_i
\tr(\sigma_\alpha B_i).
\e
Moreover, the computation of the entire response tensor of
Eq.~(\ref{sup_eq:spindensity1}) in the diffusive approximation can also be
expressed via the matrix $\mathcal{F}$ as
\be
\label{RRR}
\mathcal{R}_{\alpha\beta}\!=\!\frac{\alpha_0\ep\tau}{8\hbar}\, \sum_{ij} \lt[\tr
\sigma_\alpha B_i\rt] \lt[\frac{\mathcal{F}}{1-\mathcal{F}}\rt]_{ij} \lt[\tr
\sigma_\beta B_j\rt],
\e
where $\alpha_0=\mathcal{A}\Delta_\textrm{sd}^2/\pi\hbar^2v_f^2S$ is the
coefficient used in Eq.~(\ref{GDD}) to define the unit of the Gilbert damping. 

It appears that one can always choose the basis of $B_i$ operators such that the
computation of Eq.~(\ref{RRR}) is closed in a subspace of just three $B_i$
operators with $i=1,2,3$. This enables us to make analytical computations of
Eq.~(\ref{RRR}).   

\section{Magnetization dynamics}

The representation of the results can be made somewhat simpler by choosing $x$
axis in the direction of the in-plane projection $\bs{n}_\para$ of the N\'eel
vector, hence $n_y=0$. In this case, one can represent the result as
\be
\delta \bs{s}^+ = c_1 \bs{n}_\para \times (\bs{n}_\para \times \partial_t
\bs{m}_\para) + c_2 \partial_t \bs{m}_\para + c_3 \partial_t \bs{m}_\perp + c_4
\bs{n},\n
\e
where $\bs{n}$ dependence of the coefficients $c_i$ may be parameterized as 
\beml
\begin{align}
c_1 & = \frac{r_{11}-r_{22}-r_{31}(1-n_z^2)/(n_x n_z)}{1-n_z^2},\\
c_2 & = r_{11}-r_{31}(1-n_z^2)/(n_x n_z),\\
c_3 & = r_{33},\\
c_4 & = (r_{31}/n_z)\,\partial_t m_z + \zeta (\partial_t \bs{m})\cdot \bs{n}.
\end{align}
\eml

The analytical results in the paper correspond to the evaluation of $\delta
\bs{s}^\pm$ up to the second order in $\Delta_\textrm{sd}$ using perturbative
analysis. Thus, zero approximation corresponds to setting $\Delta_\textrm{sd}=0$
in Eqs.~(\ref{sup:eq:Ham},\ref{sup:eq:Sigma}). 

The equations of motion on $\bs{n}$ and $\bs{m}$ are given by
Eqs.~(\ref{eq:EOM}),
\beml
\label{sup:eq:EOM}
\begin{align}
\label{sup:eq:EOMa}
\pa_t\bs{n} =\,& -J\,\bs{n}\times\bs{m} + \bs{n}\times\delta\bs{s}^+ + \bs{m}\times\delta\bs{s}^-,\\
\label{sup:eq:EOMb}
\pa_t\bs{m} =\,& \bs{m}\times\delta\bs{s}^+ + \bs{n}\times\delta\bs{s}^-,
\end{align}
\eml
It is easy to see that the following transformation leaves the above equations
invariant,
\be
\delta \bs{s}^+ \rightarrow \delta \bs{s}^+ - \xi\,\bs{n},\qquad  \delta
\bs{s}^- \rightarrow \delta \bs{s}^- - \xi\, \bs{m}, 
\e
for an arbitrary value of $\xi$. 

Such a gauge transformation can be used to prove that the coefficient $c_4$ is
irrelevant in Eqs.~(\ref{sup:eq:EOM}).

In this paper, we compute $\delta \bs{s}^\pm$ to the zeroth order in $|\bs{m}|$
-- the approximation which is justified by the sub-lattice symmetry in the
anti-ferromagnet. A somewhat more general model has been analyzed also in
Ref.~\onlinecite{baglai_giant_2020} to which we refer the interested reader for
more technical details. 

\section{Anisotropy constant}

The anisotropy constant is obtained from the grand potential energy $\Omega$ for
conducting electrons. For the model of Eq.~(\ref{sup:eq:Ham}) the latter can be
expressed as 
\be
\Omega = - \sum_{\varsigma=\pm}\frac{1}{\beta} \int \mathrm{d}\ep\,g(\ep)
\nu_\varsigma(\ep),
\label{eq:bnu}
\end{equation}
where $\beta=1/k_\textrm{B}T$ is the inverse temperature,  $\varsigma=\pm$ is
the valley index (for the valleys $\bb{K}$ and $\bb{K}'$),
$G^\text{R}_{\varsigma,\bb{p}}$ is the bare retarded Green function with
momentum $\bb{p}$ and in the valley $\varsigma$. We have also defined the
function
\be
g(\ep) = \log \lt(1+\exp[\beta(\mu-\ep)]\rt),
\e
where $\mu$ is the electron potential, and the electron density of states in
each of the valleys is given by,
\be
\nu_\varsigma(\ep) =  \frac{1}{\pi} \int
\frac{\mathrm{d}\bs{p}}{(2\pi\hbar)^2}\,\im \tr G^\text{R}_{\varsigma,\bs{p}},
\e
where the trace is taken only over spin and sub-lattice space, 

In the metal regime considered, the chemical potential is assumed to be placed
in the upper electronic band. In this case, the energy integration can be taken
only for positive energies. The two valence bands are always filled and can only
add a constant shift to the grand potential $\Omega$ that we disregard. 

The evaluation of Eq.~(\ref{eq:bnu}) yields the following density of states
\be
\nu_\tau(\ep)  = \frac{1}{2\pi \hbar^2 v_f^2}
\bc
0 & 0 < \ep < \ep_2\\
\ep/2 +\lambda/4 &  \ep_2 < \ep < \ep_1,\\
\ep & \ep > \ep_1,
\ec
\e
where the energies $\ep_{1,2}$ correspond to the extremum points (zero velocity)
for the electronic bands. These energies, for each of the valleys, are given by 
\beml
\label{eq:elimits}
\begin{align}
\ep_{1,\varsigma} & = \frac{1}{2}\big(+\lambda+\sqrt{4\Delta^2+\lambda^2-4\varsigma\Delta\lambda n_z} \big),    \label{eq:elimits1}\\
\ep_{2,\varsigma} & = \frac{1}{2}\big(-\lambda+\sqrt{4\Delta^2+\lambda^2+4\varsigma\Delta\lambda n_z} \big)
\label{eq:elimits2}
\end{align}
\eml
where $\varsigma=\pm$ is the valley index. 

In the limit of zero temperature we can approximate Eq.~(\ref{eq:bnu}) as
\be
\Omega = -\sum_{\varsigma=\pm}\frac{1}{\beta}\int_0^\infty
\mathrm{d}\ep\,(\mu-\ep)\nu_\varsigma(\ep).
\label{eq:newomega}
\e
Then, with the help of Eq.~(\ref{eq:bnu}) we find, 
\begin{align}
\Omega =\,& - \frac{1}{24\pi\hbar^2v_f^2}\sum_{\varsigma=\pm}\lt[(\ep_{1,\varsigma}-\mu)^2(4\ep_{1,\varsigma}-3\lambda+2\mu)\rt.\n\\
& \lt.+(\ep_{2,\varsigma}-\mu)^2(4\ep_{2,\varsigma}+3\lambda+2\mu)\rt].
\end{align}

By substituting the results of Eqs.~(\ref{eq:elimits}) into the above equation
we obtain
\begin{align}
\Omega =\,& - \frac{1}{24\pi\hbar^2v_f^2}\lt[ (4\Delta^2-4n_z \Delta\lambda+\lambda^2)^{2/3} \rt.\n\\
&\lt.+(4\Delta^2+4n_z \Delta\lambda+\lambda^2)^{2/3}-24\Delta\mu+8\mu^3 \rt].
\end{align}
A careful analysis shows that the minimal energy corresponds to $n_z=\pm1$ so
that the conducting electrons prefer an easy-axis magnetic anisotropy. By
expanding in powers of $n_z^2$ around $n_z=\pm1$ we obtain $\Omega = -
Kn_z^2/2$, where
\be
K= \frac{1}{2\pi\hbar^2v^2}
\bc
|\Delta^2\lambda| &  |\lambda/2\Delta| \geq 1, \\
|\Delta\lambda^2|/2 &  |\lambda/2\Delta| \leq 1.
\ec
\e
This provides us with the easy axis anisotropy of Eq.~(\ref{Kvalue}).

%

\end{document}